\documentclass{aa}
\usepackage{savesym}
\savesymbol{bibfont}
\usepackage{amsmath}
\usepackage{lscape}
\usepackage{rotating}
\usepackage{natbib}
\usepackage[normalem]{ulem}
\usepackage{graphicx}
\usepackage{txfonts}
\setcitestyle{aysep={}}
\restoresymbol{NAT}{bibfont}

\usepackage{natbib,twoopt}
\usepackage[breaklinks=true]{hyperref} %% to avoid \citeads line fills
\bibpunct{(}{)}{;}{a}{}{,}             %% natbib format for A&A and ApJ

\begin{document}

\title{Limits on quantum gravity effects from \textit{Swift} short gamma-ray bursts}

\author{M. G. Bernardini\inst{1,2}\thanks{E--mail:bernardini@lupm.in2p3.fr} \and G. Ghirlanda\inst{2} \and S. Campana\inst{2} \and P. D'Avanzo\inst{2} \and J.-L. Atteia\inst{3,4} \and S. Covino\inst{2} \and G. Ghisellini\inst{2} \and A. Melandri\inst{2} \and F. Piron\inst{1} \and R. Salvaterra\inst{5} \and G. Tagliaferri\inst{2}}

\institute{Laboratoire Univers et Particules de Montpellier, Universit\'e de Montpellier, CNRS/IN2P3, Montpellier, France \and INAF--Osservatorio Astronomico di Brera, via E. Bianchi 46, I--23807 Merate, Italy \and Universit\'e de Toulouse, UPS-OMP, IRAP, Toulouse, France \and CNRS IRAP, 14 avenue Edouard Belin, F-31400 Toulouse, France \and INAF-IASF Milano, via E. Bassini 15, I-20133 Milano, Italy}

\titlerunning{Limits on quantum gravity effects from \textit{Swift} S-GRBs}

\authorrunning{Bernardini et al.}

\date{}

\abstract{The delay in arrival times between high and low energy photons from cosmic sources can be used to test the violation of the Lorentz invariance (LIV), predicted by some quantum gravity theories, and to constrain its characteristic energy scale ${\rm E_{QG}}$ that is of the order of the Planck energy. Gamma-ray bursts (GRBs) and blazars are ideal for this purpose thanks to their broad spectral energy distribution and cosmological distances: at first order approximation, the constraints on ${\rm E_{QG}}$ are proportional to the photon energy separation and the distance of the source. However, the LIV tiny contribution to the total time delay can be dominated by intrinsic delays related to the physics of the sources: long GRBs typically show a delay between high and low energy photons related to their spectral evolution (spectral lag). Short GRBs have null intrinsic spectral lags and are therefore an ideal tool to measure any LIV effect. We considered a sample of $15$ short GRBs with known redshift observed by \emph{Swift} and we estimate a limit on ${\rm E_{QG}}\gtrsim 1.5\times 10^{16}$ GeV. Our estimate represents an improvement with respect to the limit obtained with a larger (double) sample of long GRBs and is more robust than the estimates on single events because it accounts for the intrinsic delay in a statistical sense.}


\maketitle

\section{Introduction}\label{sect_i}

A quantum theory of gravity is expected to reconcile the classical theory of gravity and quantum physics. An unanimous quantum theory of gravity does not, however, exist yet. In general such theories predict the existence of a natural scale at which Einstein's classical theory breaks down. This is the quantum gravity energy scale $E_{\rm QG}$, expected to be of the order of the Planck energy $E_{\rm p}=\sqrt{(\hbar c^5)/G}\sim 1.22\times 10^{19}$ GeV. Some approaches to quantum gravity predict a deformation of the dispersion law of photons, $(c{\bf p})^2=E^2\left[1+f(E/E_{\rm QG})\right]$, where {\bf p} is the photon momentum and $c$ is the velocity of light, that would lead to energy-dependent velocities for massless particles \citep{1998Natur.393..763A,2005LRR.....8....5M}. At smaller energies, a series expansion can be applicable, that at first order would lead to an energy-dependent photon velocity $v$ of the form:
\begin{equation}
\frac{v}{c} \approx \left(1-\xi \frac{E}{E_{\rm QG}}\right)\, ,
\end{equation}
where $\xi=\pm 1$ is a sign ambiguity that can be fixed in a specific quantum gravity theory. In what follows we assume that the sign of the effect does not depend on the photon polarisation, that is, the velocities of all photons of the same energy are either increased or decreased by the same exact amount. The dependence of $\xi$ on the polarisation produces a frequency-dependent rotation of the polarisation vector in linearly polarised light, known as vacuum birefringence \citep{2005LRR.....8....5M}. 

An energy-dependent speed of photons would imply that two photons emitted simultaneously with energies $E_1$ and $E_2$ traveling a distance $L$ accumulate a delay $\Delta t_{\rm QG}$:
\begin{equation}
\Delta t_{\rm QG} \approx \xi \frac{(E_2-E_1)}{E_{\rm QG}} \frac{L}{c} \sim 10^{-2} \left(\frac{E_{\rm p}}{E_{\rm QG}}\right) \left(\frac{E_2-E_1}{\rm MeV}\right) \left(\frac{L}{\rm Mpc}\right)\, {\rm ms}\, .\label{deltae}
\end{equation}
There are two ways to magnify this delay in order to measure it: i) to increase the separation between $E_1$ and $E_2$ and/or ii) search for this effect in sources at cosmological distances. 

\begin{figure*}
\centering
\includegraphics[width= \hsize,clip]{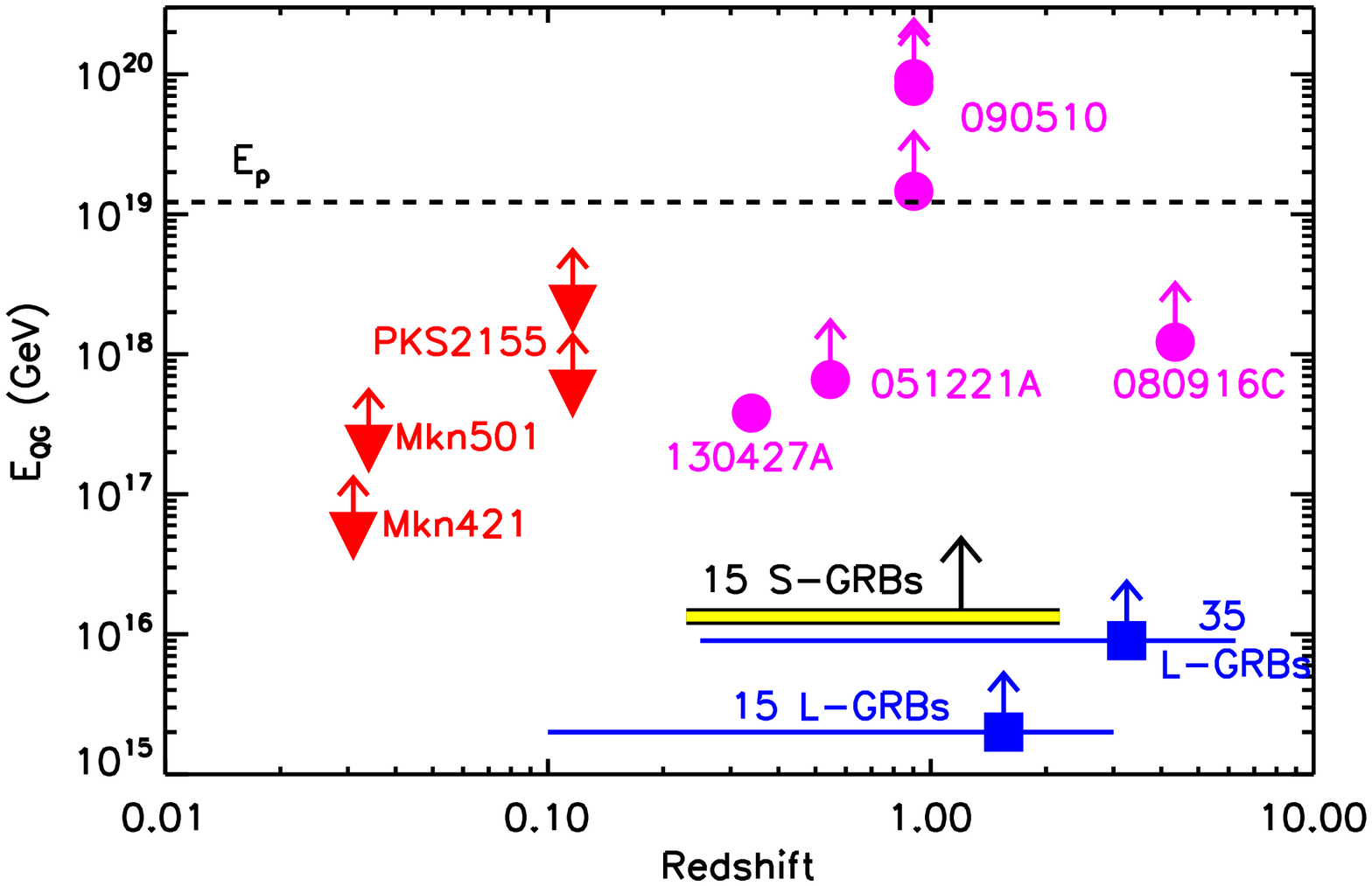}
\caption{Current limits on the quantum gravity energy scale available in the literature from extragalactic sources: TeV blazars (red triangles: Mkn 421, \citealt{1999PhRvL..83.2108B}; Mkn 501, \citealt{2008PhLB..668..253M}; PKS 2155-304, \citealt{2008PhRvL.101q0402A,2011APh....34..738H}), single GRBs (magenta points: S-GRB 051221A, \citealt{2006JCAP...05..017R}; L-GRB 080916C, \citealt{2009Sci...323.1688A}; S-GRB 090510, \citealt{2009Natur.462..331A,2010A&A...510L...7G,2013PhRvD..87l2001V}; L-GRB 130427A, \citealt{2013arXiv1305.2626A}) and samples of L-GRBs (blue squares, \citealt{2008ApJ...676..532B,2008APh....29..158E}), compared to the result obtained in the present work.}
\label{fig_lim}
\end{figure*}

Gamma-ray bursts (GRBs) are exquisite to this purpose: they are observed at cosmological distances (up to redshift 9.2) and their emission during the prompt phase can span several orders of magnitude in energy. The bright short GRB 051221A has been used to set a stringent constraint to $E_{\rm QG}$ using Konus-Wind data: $E_{\rm QG}> 0.1\,E_{\rm p}$ (\citealt{2006JCAP...05..017R}, see also \citealt{2006JCAP...04..006R} for a discussion about the methodology). The \emph{Fermi} gamma-ray telescope, with its broad energy range (from few keV to several GeV), has enabled us to test possible violations of the Lorentz invariance studying the delays among different energy bands, maximising the energy difference. \citet{2009Sci...323.1688A} used the long GRB 08091C and its highest energy detected photon ($13.2$ GeV) to estimate the maximum delay ($16.5$ s after the trigger) and in turn to provide a constrain: $E_{\rm QG}> 0.1\,E_{\rm p}$. For the short GRB 090510, the time delay between the trigger time and the arrival time of one $31$ GeV photon was estimated to be $0.86$ s. This led \citet{2009Natur.462..331A} to set a stringent limit on $E_{\rm QG} > 1.2\,E_{\rm p}$. \citet{2010A&A...510L...7G} and \citet{2013PhRvD..87l2001V} obtained even tighter constraints for the same GRB with different assumptions ($E_{\rm QG}>6.7\,E_{\rm p}$ and $E_{\rm QG}>7.6\,E_{\rm p}$, respectively). Figure~\ref{fig_lim} portrays the limits currently derived with GRBs and other extragalactic sources. Measures of GRB polarisation have been used to constrain the vacuum birefringence effect \citep{2007MNRAS.376.1857F,2011PhRvD..83l1301L,2012PhRvL.109x1104T,2013MNRAS.431.3550G,2014MNRAS.444.2776G,2016MNRAS.463..375L}.

%Single events however might be plagued by modelling uncertainties, so that statistical samples have to be preferred. 
\citet{2008ApJ...676..532B} and \citet{2008APh....29..158E} performed a statistical study on samples of long GRBs (L-GRBs) with redshift detected by several instruments (HETE-2, BATSE and \textit{Swift}), deriving that the energy scale $E_{\rm QG}>2\times 10^{15}$ GeV and $E_{\rm QG}>9\times 10^{15}$ GeV, respectively (see Fig.~\ref{fig_lim}). Though these limits are less stringent than the ones obtained with single events, this statistical approach has the merit to help in disentangling the quantum gravity effect from some intrinsic effects, as for example the spectral lag.

A delay of high and low energy photons or spectral lag is a well-known property of GRBs \citep{1995A&A...300..746C,2000ApJ...534..248N,2002ApJ...579..386N,2006Natur.444.1044G,2007ApJ...660...16S,2008ApJ...677L..81H,2010PASJ...62..487A,2010ApJ...711.1073U}. Short GRBs (S-GRBs) are consistent with negligible spectral lag \citep{2001grba.conf...40N,2006ApJ...643..266N,2015MNRAS.446.1129B}, while L-GRBs have positive, null or negative lag \citep{2012MNRAS.419..614U,2015MNRAS.446.1129B}. 
%Even when the spectral lag is negligible, it is way larger than the expected delay induced by quantum gravity effects. Besides, the spectral lag distribution is quite broad for both L- and S-GRBs, being spread over two and one orders of magnitude, respectively.
No clear physical motivation helps one in accounting for the spectral lag contribution in each single case (for possible interpretations of the spectral lag see \citealt{1998ApJ...501L.157D,2003ApJ...594..385K,2005A&A...429..869R,2011AN....332...92P,2000ApJ...544L.115S,2001ApJ...554L.163I,2004ApJ...614..284D,2005MNRAS.362...59S,2006MNRAS.367..275L,2010ApJ...725..225G,2013ApJ...770...32G,2016A&A...592A..95M}). Thus, when deriving the limits for quantum gravity effects in specific GRBs, either long or short, we cannot properly model the contribution of the intrinsic spectral lag to estimate the possible delay induced by quantum gravity effects alone (see however \citealt{2013PhRvD..87l2001V}, who accounted for the intrinsic spectral lag in a statistical sense in single sources). 

%For S-GRBs the typical spectral lag is null or consistent with zero. Therefore, adopting a large sample of SGRBs ensures us that the spectral lag contribution to the average delay cancels out, so that any measurement of the delay is more likely due or interpretable as arising from dispersion of the velocity of photons. 

%----------------------------------
%----------------------------------

In this paper we have adopted a statistical approach as in \citet{2008ApJ...676..532B} and \citet{2008APh....29..158E} to single out properly the source contribution to the delay. At variance with previous studies, we considered the largest possible sample of S-GRBs observed by the \textit{Swift}/Burst Alert Telescope (BAT; \citealt{2005SSRv..120..143B}). This sample provides three main advantages: i) the spectral lag of S-GRBs is negligible; ii) the dispersion of the spectral lag of S-GRBs is much smaller than for L-GRBs \citep{2015MNRAS.446.1129B}; iii) the use of a single instrument reduces also the possible systematics that arise when combining data from different instruments \citep{2008APh....29..158E}; iv) \textit{Swift}/BAT enables us to perform the analysis on the largest available sample of S-GRBs with redshift.

In Section~\ref{sample} we describe the sample selection and the methodology used to derive the time delay. In Section~\ref{limits} we detail the derivation of the limit on the quantum gravity energy scale. In Section~\ref{conc} we discuss our results. Errors are given at $1\,\sigma$ confidence level, unless otherwise stated. We used the cosmological parameters based on full-mission Planck observations \citep{2016A&A...594A..13P}.

\section{Sample selection and methodology}\label{sample}

We selected the \textit{Swift} GRBs classified as short by the BAT team refined analysis, namely all the GRBs with $T_{90} < 2$ s and those whose \textit{Swift}/BAT light curve shows a short-duration peak followed by a softer, long-lasting tail (the so-called extended emission, with $T_{90} > 2$ s). We also required that these GRBs have a redshift measurement\footnote{We excluded GRB 080905A whose redshift has been questioned in \citet{2014arXiv1405.5131D}.}. We excluded from our analysis GRB 090426 and GRB 100816A since \citet{2014arXiv1405.5131D} considered them as possible L-GRBs (i.e. they likely have a collapsar progenitor, \citealt{2006Natur.444.1010Z,2013ApJ...764..179B}). We ended up with $21$ S-GRBs with redshift. Most of these events and their prompt emission properties are reported in \citet{2014arXiv1405.5131D}.

In order to calculate the time delay $\Delta t$ between photons of high and low energy, we exploited the same methodology adopted in \citet{2015MNRAS.446.1129B} for the calculation of the spectral lag, namely:
\begin{itemize}
\item we extracted mask-weighted, background-subtracted light curves with the \texttt{batmaskwtevt} and \texttt{batbinevt} tasks in FTOOLS for two fixed observer frame energy bands (ch1: $50-100$ keV and ch2: $150-200$ keV) within the energy range of the BAT instrument ($\sim [15-200]$ keV; \citealt{2011ApJS..195....2S});
\item we used the discrete cross-correlation function (CCF; \citealt{1997ApJ...486..928B}) to measure the temporal correlation of the two light curves in ch1 and ch2. We calculated the CCF value for a series of time delays over the entire light curve that are multiples of the time resolution of the light curves. The temporal delay of the photons is defined as the global maximum of the CCF. For each GRB we tried different time resolutions and we used the minimum one with a chance probability $< 10^{-3}$ of finding the corresponding CCF$_{\rm max}$ to discard statistical fluctuations;
\item to locate the global maximum, we fitted an asymmetric Gaussian model to the CCF. This allows us to estimate lags which can be a fraction of the time resolution of the light curves extracted from the BAT data. The uncertainties on the CCF and on the time delay have been derived by applying a flux-randomisation method \citep{1998PASP..110..660P}, as described in \citet{2015MNRAS.446.1129B}.
\end{itemize}
We applied this procedure to the $21$ short GRBs of our sample and ended up with $15$ short GRBs with a significative global maximum in the CCF. These GRBs spaned a redshift range $z\in[0.36-2.2]$. Six GRBs (GRB 050724, GRB 070724, GRB 080123, GRB 131004A, GRB 140903A and GRB 150120A) have been discarded from our analysis because there were not enough counts in one of the two energy bands to extract a significant value of the CCF for any choice of the temporal resolution. The results of the analysis are reported in Table~\ref{tab1}.

%----------------------------------
\begin{figure*}
\centering
\includegraphics[width= \hsize,clip]{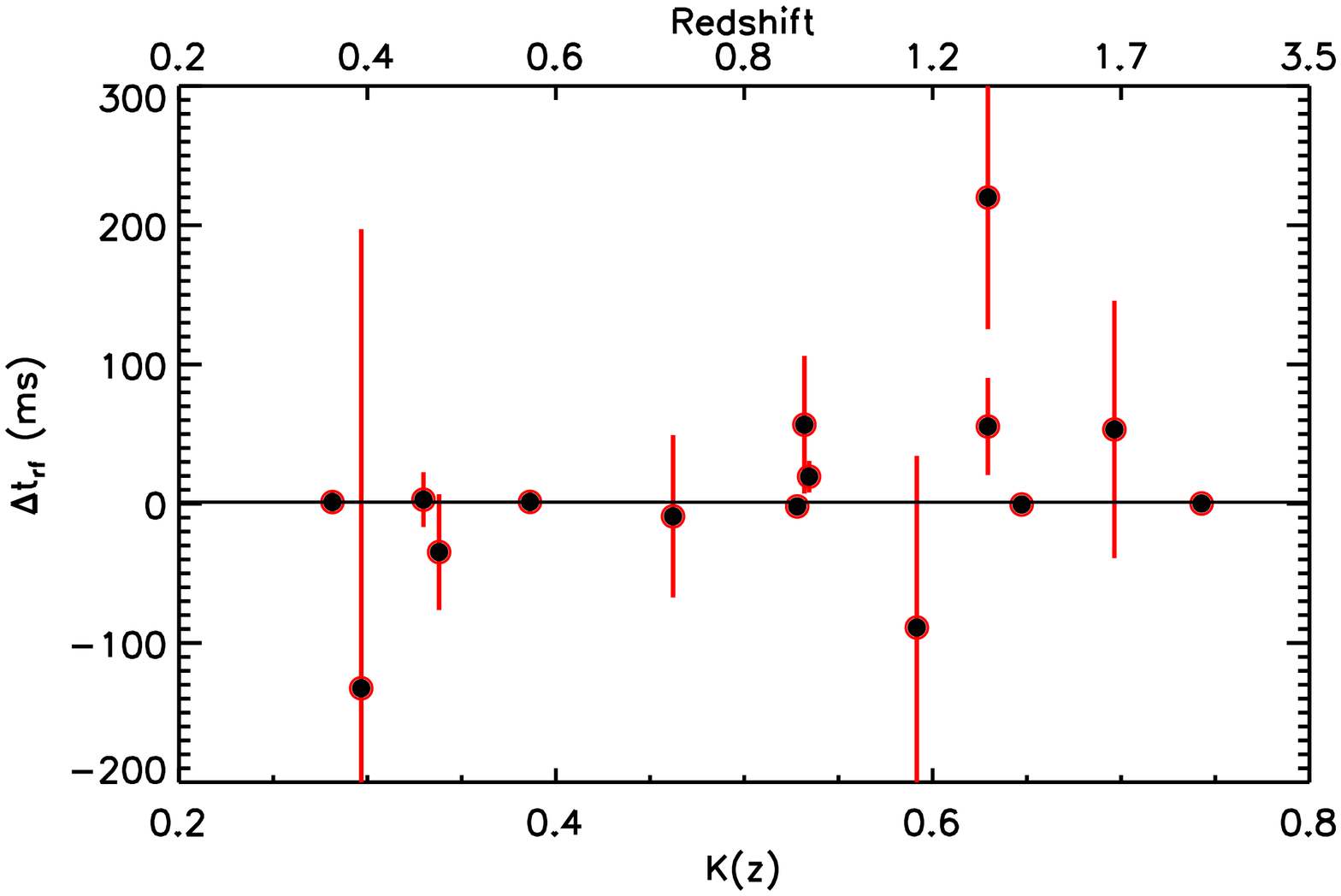}
\caption{Temporal delay $\Delta t_{\rm rf}$ as a function of the distance term $K(z)$ for the $15$ S-GRBs of our sample. The black line marks the best fit: $\Delta t_{\rm rf}/$ms$= 0.95+0.11\,K(z)$. The redshift scale is reported in the upper axis.}
\label{fig_del_k}
\end{figure*}
%----------------------------------

\begin{table*}
\caption{Spectral lags for the $15$ S-GRBs of our samples. GRB name, redshift ($z$), distance term $K(z)$, temporal resolution (bin), left ($t_l$), and right ($t_r$) boundaries of the time interval over which the spectral lag is computed, spectral lag in the observer frame ($\tau$), left ($\sigma_l$) and right ($\sigma_r$) uncertainties.}
\label{tab1} 
\begin{center}
\begin{tabular}{ccccccccc}
\hline
\hline
GRB name & $z$ & $K(z)$ & bin (ms) & $t_l$ (s) & $t_r$ (s) & $\tau$ (ms) & $\sigma_l$ (ms) & $\sigma_r$ (ms)\\
\hline                                                                                                                                                                 
051210 & $     1.30^{*a}$ & $     0.63$ & $  64$ & $     -0.5$ & $      1.8$ & $   505.87$ & $   197.59$ & $   237.40$ \\
051221A & $     0.55^a$ & $     0.39$ & $   8$ & $     -0.3$ & $      0.5$ & $     2.01$ & $    12.37$ & $    13.06$ \\
060801 & $     1.13^a$ & $     0.59$ & $ 128$ & $      0.0$ & $      3.0$ & $  -189.38$ & $   249.96$ & $   275.24$ \\
061006 & $     0.44^a$ & $     0.33$ & $  32$ & $    -22.9$ & $    -22.2$ & $     4.19$ & $    27.10$ & $    29.53$ \\
070714B & $     0.92^a$ & $     0.53$ & $  32$ & $     -1.0$ & $      2.0$ & $    37.36$ & $    22.46$ & $    21.19$ \\
071227 & $     0.38^a$ & $     0.30$ & $ 256$ & $     -1.2$ & $      2.2$ & $  -183.03$ & $   471.46$ & $   439.03$ \\
090510 & $     0.90^a$ & $     0.53$ & $  16$ & $     -0.2$ & $      0.5$ & $    -3.71$ & $     9.11$ & $     9.23$ \\
100117A & $     0.92^a$ & $     0.53$ & $ 128$ & $     -1.0$ & $      0.8$ & $   108.64$ & $    91.63$ & $    97.54$ \\
100625A & $     0.45^a$ & $     0.34$ & $  32$ & $     -0.2$ & $      0.5$ & $   -50.47$ & $    64.53$ & $    56.21$ \\
101219A & $     0.72^a$ & $     0.46$ & $  64$ & $     -0.5$ & $      1.0$ & $   -15.56$ & $   111.80$ & $    88.36$ \\
111117A & $     2.20^b$ & $     0.74$ & $  16$ & $     -0.5$ & $      1.0$ & $     0.56$ & $    12.53$ & $    12.54$ \\
120804A & $     1.30^{*a}$ & $     0.63$ & $  32$ & $     -1.0$ & $      1.0$ & $   127.54$ & $    79.54$ & $    80.64$ \\
130603B & $     0.36^a$ & $     0.28$ & $   8$ & $     -0.3$ & $      0.3$ & $     1.65$ & $     9.65$ & $    11.40$ \\
150423A & $     1.39^c$ & $     0.65$ & $  16$ & $     -0.1$ & $      0.5$ & $    -1.32$ & $    13.64$ & $    13.50$ \\
160410A & $     1.72^d$ & $     0.70$ & $ 128$ & $     -1.0$ & $      2.0$ & $   145.06$ & $   264.46$ & $   237.85$ \\
\hline
\multicolumn{2}{l}{$^*$ photometric redshift}\\
\multicolumn{2}{l}{$^a$ \citet{2014arXiv1405.5131D}}\\
\multicolumn{2}{l}{$^b$ \citet{2017arXiv170701452S}}\\
\multicolumn{2}{l}{$^c$ \citet{2015GCN..17755...1M}}\\
\multicolumn{2}{l}{$^d$ \citet{2016GCN..19274...1S}}
\end{tabular}
\end{center}
\end{table*}

\section{Limits to the quantum-gravity energy scale from S-GRBs}\label{limits}

The temporal delay between high and low energy photons can be written as
\begin{equation}
\Delta t = \tau + \Delta t^{\rm QG}\,
,\end{equation}
where $\tau$ is the contribution to the delay intrinsic to the GRB (the intrinsic spectral lag), while $\Delta t^{\rm QG}$ is the systematic delay induced by the violation of Lorentz invariance. The second term corresponds to the delay in arrival time of two photons with energy difference $\Delta E$ in the observer frame, emitted simultaneously by a cosmological source located at redshift $z$ (see \citealt{2008JCAP...01..031J} for a complete derivation of this formula):
\begin{equation}
\Delta t^{\rm QG} = H_\circ^{-1}\frac{\Delta E}{E_{\rm QG}} \int_0^z \frac{(1+z') dz'}{h(z')}\, ,
\label{eq_tqg_1}
\end{equation}
where $H_\circ$ is the Hubble expansion rate and $h(z')=\sqrt{\Omega_{\Lambda}+\Omega_M (1+z')^3}$. For convenience, we rewrite Eq.~\ref{eq_tqg_1} in terms of the time delay measured in the source rest frame:
\begin{equation}
\Delta t_{\rm rf}^{\rm QG} = \frac{\Delta t^{\rm QG}}{(1+z)} = H_\circ^{-1}\frac{\Delta E}{E_{\rm QG}} \left[ \frac{1}{(1+z)} \int_0^z \frac{(1+z') dz'}{h(z')}\right]\, .
\label{eq_tqg}
\end{equation}
Overall, the time delay can be written as a linear function:
\begin{equation}
\Delta t_{\rm rf} = q_{\rm s} + m_{\rm QG} K(z)\, ,
\label{eq_fit}
\end{equation}
where $K(z)$ contains the dependence of the temporal delay upon the distance (the quantity in square brackets in Eq.~\ref{eq_tqg}) and $q_{\rm s}=\tau/(1+z)$.

We computed for each S-GRB of our sample the corresponding $K(z)$ (see Table~\ref{tab1}) and fitted to the data the model in Eq.~\ref{eq_fit} to determine the coefficient $m_{\rm QG}$. The intercept $q_{\rm s}=\tau_{\rm rf}$ represents the contribution from the intrinsic spectral lag. The rest-frame temporal delay as a function of $K(z)$ is portrayed in Fig.~\ref{fig_del_k}. We considered an extra-scatter $\sigma_{\rm s}$ that accounts for the dispersion of the intrinsic spectral lag. Markov chain Monte Carlo techniques are used in our calculations to derive the best-fitting parameters: for each Markov chain, we generated $10^5$ samples according to the likelihood function\footnote{In our analysis we used JAGS (Just Another Gibbs Sampler).  It is a programme for analysis of Bayesian hierarchical models using Markov chain Monte Carlo simulation. More information can be found: http://mcmc-jags.sourceforge.net/}. Then we derived coefficients and confidence interval according to the statistical results of the samples. This yields: 
%$\Delta t_{\rm rf} = (1.14_{-3.49}^{+3.47})$ ms $+ [(-0.07_{-1.58}^{+1.56})$ ms$]\,K(z)$. The intrinsic scatter is $\sigma_{\rm s}=(4.09_{-3.22}^{+3.17})$ ms.
$\Delta t_{\rm rf} = (0.95_{-3.55}^{+3.70})$ ms $+ [(0.11_{-1.74}^{+1.54})$ ms$]\,K(z)$. The intrinsic scatter is $\sigma_{\rm s}=(4.18_{-3.20}^{+3.11})$ ms.

The coefficient $m_{\rm QG}$ is consistent with it being zero within $1\,\sigma$. This allows us to place a lower limit to the effective energy scale for the rising of the quantum-gravity effect, adopting the same technique described above to derive the best-fitting parameters and considering as a prior that the energy is a positive quantity: 
%$E_{\rm QG} > 1.5\times 10^{16}$ GeV ($95\%$ c.l.). %The limit is not substantially imrpoved even if we assume that S-GRBs have intrinsically null spectral lag ($q_s=0$) and perform the same analysis.
$E_{\rm QG} > 1.48\times 10^{16}$ GeV ($95\%$ c.l.).

\section{Discussion and conclusions}\label{conc}

The systematic analysis of the temporal lag for S-GRBs observed by \textit{Swift} allowed us to derive a lower limit for the effective energy scale for the onset of the quantum-gravity delay. Our result is more stringent than those obtained with larger samples of L-GRBs, and is more robust than the estimates on single events because:
\begin{itemize}
\item the physical origin of the intrinsic spectral lag is still unclear, and it is not possible to predict theoretically its value for specific events. Furthermore, the intrinsic lag may be negligible, positive or negative without any apparent relation with the GRB properties \citep{2012MNRAS.419..614U,2015MNRAS.446.1129B}, thus it is hard to disentangle its contribution from the purely quantum-gravity delay of photons. \citet{2013PhRvD..87l2001V} made an attempt to account for intrinsic effects on single bright GRBs observed at GeV energies, finding strong constraints ($E_{\rm QG} > 1.8\,E_{\rm p}$ on S-GRB 090510). However, using a sample of GRBs characterised by a short single event is the best way to account for the intrinsic lag in a statistical sense (see also \citealt{2008ApJ...676..532B} and \citealt{2008APh....29..158E});
\item S-GRBs have intrinsic lag consistent with zero, with much smaller dispersion compared to L-GRBs \citep{2015MNRAS.446.1129B}. Thus, using S-GRBs we reduce the uncertainties about the intrinsic lag and its scatter, allowing us to derive more robust constraints than in similar analysis with L-GRBs, though the sample is limited in number. Adding L-GRBs with negligible intrinsic lag would not improve our estimates because though the two samples are likely drawn from the same population;
%(probability $p=16\%$\footnote{In order to account for the uncertainties on the spectral lag, that in some cases may be large, we calculated the moments of the distribution from $10,000$ distributions of spectral lags of L-GRBs with null spectral lag and S-GRBs from the original samples from \citet{2015MNRAS.446.1129B} by assuming that each spectral lag is normally distributed around the calculated value, with a standard deviation equal to its uncertainty. We performed a Kolmogorov-Smirnov (KS) test for each of the $10,000$ distributions of spectral lags, and calculated the mean probability that the two samples are drawn from the same population. 
\footnote{We caution, however, that the samples in \citet{2015MNRAS.446.1129B} are analysed in two fixed rest-frame energy bands different from ch1 and ch2 adopted in the present work. Being the spectral lag dependent upon the energy bands adopted, these figures might be slightly modified when calculated for ch1 and ch2.} \citep{2015MNRAS.446.1129B}, the dispersion for L-GRBs with null lag is much larger ($\sigma_{\rm L-GRBs}=(110\pm32)$ ms). The selection of events observed by a single instrument reduces also the possible systematics that arise when combining data from different instruments \citep{2008APh....29..158E};
\item the intrinsic spectral lag within a single GRB may evolve with time, and the time-integrated quantity parametrised by $q_s$ is only an `average' (in a non-statistical sense) representation of it. Uncertainties much larger than the temporal resolution of the light curves on the lag may be related to the convolution of multiple peaks in the CCF that spread the absolute maximum \citep{2015MNRAS.446.1129B}. This effect is much more relevant for L-GRBs than for S-GRBs;
\item S-GRBs have usually lower redshifts than long GRBs (see e.g. \citealt{2014arXiv1405.5131D}). The average redshift for L-GRBs is $\sim 1.8$, \citealt{2012ApJ...749...68S}, extending up to $z\sim 9$, \citealt{2011ApJ...736....7C}). However, the term $K(z)$ weakly increases for large redshifts ($\sim 10\%$ when passing from redshift 2 to redshift 4). Therefore, the low redshift range covered by short GRBs does not disfavour their use to probe LIV.
\end{itemize}

To evaluate if the present result is strongly dependent on the size of the sample considered, we performed a Monte Carlo simulation to evaluate how the constraint improves with the sample size. Starting from our $15$ S-GRBs, we added S-GRBs extracted randomly from the population synthesis code for S-GRBs \citep{2016A&A...594A..84G}, generating an hybrid sample of $15+45$ S-GRBs. We assigned to each synthetic S-GRBs a lag randomly extracted from the distribution of the 15 real events. The error on the lag depends on the binning size that, in turn, is chosen to have an appropriate signal to noise ratio for the lag computation. This implies that the smaller errors are for brighter bursts. For this reason, the errors on the lags are estimated from an empirical relation with the peak flux derived for the GRBs of our sample (${\rm Log}[\sigma_{\rm lag}/{\rm ms}] = 2.5-0.9 {\rm Log}[f_{\rm pk}/{\rm (ph/cm^2/s)}]$). This sample has been analysed with the same procedure described above deriving a limit on $E_{\rm QG}$. This Monte Carlo procedure has been repeated $10^5$ times averaging the corresponding $E_{\rm QG}$ estimate. We obtained an improvement of at most a factor of two in the estimate of $E_{\rm QG}$. 

Independently of sample size, a sample of events with a widest redshift range can provide better constraints on the $E_{\rm QG}$ value (see Eq. 2). The sample considered in this paper extends up to $z = 2.2$ \citep{2017arXiv170701452S}. Based on the S-GRB redshift distribution reported in \citet{2014arXiv1405.5131D} and \citet{2016A&A...594A..84G}, we expect $5\%-30\%$ of the {\it Swift} S-GRB to have $z > 2$. Considering the {\it Swift} S-GRB detection rate ($\sim 8$ yr$^{-1}$) and the efficiency in measuring their redshift (almost 3/4 of the {\it Swift} S-GRB is missing a secure redshift measurement), this translates into about one to six events with measured $z > 2$ over ten further years of {\it Swift} activity. 

In light of the above considerations, better perspectives to derive more stringent limits with S-GRBs, with all the advantages described above, could rely on the extension of the calculation of the time delay to higher energies, exploiting the GRB broad spectral energy distribution. The method proposed in this paper applied to GeV photons (i.e. a factor $10^4$ in the $(E_2-E_1)$ term of Eq.~\ref{deltae}) would give a substantial improvement in the constraint. However, there is only one S-GRBs with known redshift and GeV detection by \emph{Fermi}/LAT (GRB 090510).

\begin{acknowledgements}
The authors thank the anonymous referee for his/her useful comments. The authors acknowledge the Italian Space Agency (ASI) for financial support through the ASI-INAF contract I/004/11/1. MGB acknowledges the support of the OCEVU Labex (ANR-11-LABX-0060) and the A*MIDEX project (ANR-11-IDEX-0001-02) funded by the `Investissements d'Avenir' French government programme managed by the ANR. 
\end{acknowledgements}

\label{lastpage}


\begin{thebibliography}{60}
\expandafter\ifx\csname natexlab\endcsname\relax\def\natexlab#1{#1}\fi

\bibitem[{{Abdo} {et~al.}(2009{\natexlab{a}}){Abdo}, {Ackermann}, {Ajello},
  {Asano}, {Atwood}, {Axelsson}, {Baldini}, {Ballet}, {Barbiellini}, {Baring},
  \& et~al.}]{2009Natur.462..331A}
{Abdo}, A.~A., {Ackermann}, M., {Ajello}, M., {et~al.} 2009{\natexlab{a}},
  Nature, 462, 331

\bibitem[{{Abdo} {et~al.}(2009{\natexlab{b}}){Abdo}, {Ackermann}, {Arimoto},
  {Asano}, {Atwood}, {Axelsson}, {Baldini}, {Ballet}, {Band}, {Barbiellini}, \&
  et~al.}]{2009Sci...323.1688A}
{Abdo}, A.~A., {Ackermann}, M., {Arimoto}, M., {et~al.} 2009{\natexlab{b}},
  Science, 323, 1688

\bibitem[{{Aharonian} {et~al.}(2008){Aharonian}, {Akhperjanian}, {Barres de
  Almeida}, {Bazer-Bachi}, {Becherini}, {Behera}, {Beilicke}, {Benbow},
  {Bernl{\"o}hr}, {Boisson}, {Bochow}, {Borrel}, {Braun}, {Brion}, {Brucker},
  {Brun}, {B{\"u}hler}, {Bulik}, {B{\"u}sching}, {Boutelier}, {Carrigan},
  {Chadwick}, {Charbonnier}, {Chaves}, {Chounet}, {Clapson}, {Coignet},
  {Costamante}, {Dalton}, {Degrange}, {Deil}, {Dickinson}, {Djannati-Ata{\"i}},
  {Domainko}, {Drury}, {Dubois}, {Dubus}, {Dyks}, {Egberts}, {Emmanoulopoulos},
  {Espigat}, {Farnier}, {Feinstein}, {Fiasson}, {F{\"o}rster}, {Fontaine},
  {F{\"u}{\ss}ling}, {Gabici}, {Gallant}, {G{\'e}rard}, {Giebels},
  {Glicenstein}, {Gl{\"u}ck}, {Goret}, {Hadjichristidis}, {Hauser}, {Hauser},
  {Heinz}, {Heinzelmann}, {Henri}, {Hermann}, {Hinton}, {Hoffmann}, {Hofmann},
  {Holleran}, {Hoppe}, {Horns}, {Jacholkowska}, {de Jager}, {Jung},
  {Katarzy{\'n}ski}, {Kaufmann}, {Kendziorra}, {Kerschhaggl}, {Khangulyan},
  {Kh{\'e}lifi}, {Keogh}, {Komin}, {Kosack}, {Lamanna}, {Lenain}, {Lohse},
  {Marandon}, {Martin}, {Martineau-Huynh}, {Marcowith}, {Maurin}, {McComb},
  {Medina}, {Moderski}, {Moulin}, {Naumann-Godo}, {de Naurois}, {Nedbal},
  {Nekrassov}, {Niemiec}, {Nolan}, {Ohm}, {Olive}, {de O{\~n}a Wilhelmi},
  {Orford}, {Osborne}, {Ostrowski}, {Panter}, {Pedaletti}, {Pelletier},
  {Petrucci}, {Pita}, {P{\"u}hlhofer}, {Punch}, {Quirrenbach}, {Raubenheimer},
  {Raue}, {Rayner}, {Renaud}, {Rieger}, {Ripken}, {Rob}, {Rosier-Lees},
  {Rowell}, {Rudak}, {Ruppel}, {Sahakian}, {Santangelo}, {Schlickeiser},
  {Sch{\"o}ck}, {Schr{\"o}der}, {Schwanke}, {Schwarzburg}, {Schwemmer},
  {Shalchi}, {Skilton}, {Sol}, {Spangler}, {Stawarz}, {Steenkamp}, {Stegmann},
  {Superina}, {Tam}, {Tavernet}, {Terrier}, {Tibolla}, {van Eldik},
  {Vasileiadis}, {Venter}, {Vialle}, {Vincent}, {Vivier}, {V{\"o}lk}, {Volpe},
  {Wagner}, {Ward}, {Zdziarski}, {Zech}, \&
  {H.E.S.S.~Collaboration}}]{2008PhRvL.101q0402A}
{Aharonian}, F., {Akhperjanian}, A.~G., {Barres de Almeida}, U., {et~al.} 2008,
  Physical Review Letters, 101, 170402

\bibitem[{{Amelino-Camelia} {et~al.}(1998){Amelino-Camelia}, {Ellis},
  {Mavromatos}, {Nanopoulos}, \& {Sarkar}}]{1998Natur.393..763A}
{Amelino-Camelia}, G., {Ellis}, J., {Mavromatos}, N.~E., {Nanopoulos}, D.~V.,
  \& {Sarkar}, S. 1998, Nature, 393, 763

\bibitem[{{Amelino-Camelia} {et~al.}(2013){Amelino-Camelia}, {Fiore}, {Guetta},
  \& {Puccetti}}]{2013arXiv1305.2626A}
{Amelino-Camelia}, G., {Fiore}, F., {Guetta}, D., \& {Puccetti}, S. 2013, ArXiv
  1305.2626 [\eprint[arXiv]{1305.2626}]

\bibitem[{{Arimoto} {et~al.}(2010){Arimoto}, {Kawai}, {Asano}, {Hurley},
  {Suzuki}, {Nakagawa}, {Shimokawabe}, {Pazmino}, {Sato}, {Matsuoka},
  {Yoshida}, {Tamagawa}, {Shirasaki}, {Sugita}, {Takahashi}, {Atteia},
  {Pelangeon}, {Vanderspek}, {Graziani}, {Prigozhin}, {Villasenor}, {Jernigan},
  {Crew}, {Sakamoto}, {Ricker}, {Woosley}, {Butler}, {Levine}, {Doty},
  {Donaghy}, {Lamb}, {Fenimore}, {Galassi}, {Boer}, {Dezalay}, {Olive},
  {Braga}, {Manchanda}, \& {Pizzichini}}]{2010PASJ...62..487A}
{Arimoto}, M., {Kawai}, N., {Asano}, K., {et~al.} 2010, Publications of the
  Astronomical Society of Japan, 62, 487

\bibitem[{{Band}(1997)}]{1997ApJ...486..928B}
{Band}, D.~L. 1997, Astrophys. J., 486, 928

\bibitem[{{Barthelmy} {et~al.}(2005){Barthelmy}, {Barbier}, {Cummings},
  {Fenimore}, {Gehrels}, {Hullinger}, {Krimm}, {Markwardt}, {Palmer},
  {Parsons}, {Sato}, {Suzuki}, {Takahashi}, {Tashiro}, \&
  {Tueller}}]{2005SSRv..120..143B}
{Barthelmy}, S.~D., {Barbier}, L.~M., {Cummings}, J.~R., {et~al.} 2005, Space
  Sci. Rev., 120, 143

\bibitem[{{Bernardini} {et~al.}(2015){Bernardini}, {Ghirlanda}, {Campana},
  {Covino}, {Salvaterra}, {Atteia}, {Burlon}, {Calderone}, {D'Avanzo},
  {D'Elia}, {Ghisellini}, {Heussaff}, {Lazzati}, {Melandri}, {Nava}, {Vergani},
  \& {Tagliaferri}}]{2015MNRAS.446.1129B}
{Bernardini}, M.~G., {Ghirlanda}, G., {Campana}, S., {et~al.} 2015, Mon. Not.
  R. Soc., 446, 1129

\bibitem[{{Biller} {et~al.}(1999){Biller}, {Breslin}, {Buckley}, {Catanese},
  {Carson}, {Carter-Lewis}, {Cawley}, {Fegan}, {Finley}, {Gaidos}, {Hillas},
  {Krennrich}, {Lamb}, {Lessard}, {Masterson}, {McEnery}, {McKernan},
  {Moriarty}, {Quinn}, {Rose}, {Samuelson}, {Sembroski}, {Skelton}, \&
  {Weekes}}]{1999PhRvL..83.2108B}
{Biller}, S.~D., {Breslin}, A.~C., {Buckley}, J., {et~al.} 1999, Phys.Rev.
  Lett., 83, 2108

\bibitem[{{Bolmont} {et~al.}(2008){Bolmont}, {Jacholkowska}, {Atteia}, {Piron},
  \& {Pizzichini}}]{2008ApJ...676..532B}
{Bolmont}, J., {Jacholkowska}, A., {Atteia}, J.-L., {Piron}, F., \&
  {Pizzichini}, G. 2008, Astrophys. J., 676, 532

\bibitem[{{Bromberg} {et~al.}(2013){Bromberg}, {Nakar}, {Piran}, \&
  {Sari}}]{2013ApJ...764..179B}
{Bromberg}, O., {Nakar}, E., {Piran}, T., \& {Sari}, R. 2013, Astrophys. J.,
  764, 179

\bibitem[{{Cheng} {et~al.}(1995){Cheng}, {Ma}, {Cheng}, {Lu}, \&
  {Zhou}}]{1995A&A...300..746C}
{Cheng}, L.~X., {Ma}, Y.~Q., {Cheng}, K.~S., {Lu}, T., \& {Zhou}, Y.~Y. 1995,
  Astron. Astrophys., 300, 746

\bibitem[{{Cucchiara} {et~al.}(2011){Cucchiara}, {Levan}, {Fox}, {Tanvir},
  {Ukwatta}, {Berger}, {Kr{\"u}hler}, {K{\"u}pc{\"u} Yolda{\c s}}, {Wu},
  {Toma}, {Greiner}, {Olivares}, {Rowlinson}, {Amati}, {Sakamoto}, {Roth},
  {Stephens}, {Fritz}, {Fynbo}, {Hjorth}, {Malesani}, {Jakobsson}, {Wiersema},
  {O'Brien}, {Soderberg}, {Foley}, {Fruchter}, {Rhoads}, {Rutledge}, {Schmidt},
  {Dopita}, {Podsiadlowski}, {Willingale}, {Wolf}, {Kulkarni}, \&
  {D'Avanzo}}]{2011ApJ...736....7C}
{Cucchiara}, A., {Levan}, A.~J., {Fox}, D.~B., {et~al.} 2011, Astrophys. J.,
  736, 7

\bibitem[{{D'Avanzo} {et~al.}(2014){D'Avanzo}, {Salvaterra}, {Bernardini},
  {Nava}, {Campana}, {Covino}, {D'Elia}, {Ghirlanda}, {Ghisellini}, {Melandri},
  {Sbarufatti}, {Vergani}, \& {Tagliaferri}}]{2014arXiv1405.5131D}
{D'Avanzo}, P., {Salvaterra}, R., {Bernardini}, M.~G., {et~al.} 2014, Mon. Not.
  R. Soc., 442, 2342

\bibitem[{{Dermer}(1998)}]{1998ApJ...501L.157D}
{Dermer}, C.~D. 1998, Astrophys. J. Lett., 501, L157

\bibitem[{{Dermer}(2004)}]{2004ApJ...614..284D}
{Dermer}, C.~D. 2004, Astrophys. J., 614, 284

\bibitem[{{Ellis} {et~al.}(2008){Ellis}, {Mavromatos}, {Nanopoulos},
  {Sakharov}, \& {Sarkisyan}}]{2008APh....29..158E}
{Ellis}, J., {Mavromatos}, N.~E., {Nanopoulos}, D.~V., {Sakharov}, A.~S., \&
  {Sarkisyan}, E.~K.~G. 2008, Astroparticle Physics, 29, 158

\bibitem[{{Fan} {et~al.}(2007){Fan}, {Wei}, \& {Xu}}]{2007MNRAS.376.1857F}
{Fan}, Y.-Z., {Wei}, D.-M., \& {Xu}, D. 2007, Mon. Not. R. Soc., 376, 1857

\bibitem[{{Gehrels} {et~al.}(2006){Gehrels}, {Norris}, {Barthelmy}, {Granot},
  {Kaneko}, {Kouveliotou}, {Markwardt}, {M{\'e}sz{\'a}ros}, {Nakar}, {Nousek},
  {O'Brien}, {Page}, {Palmer}, {Parsons}, {Roming}, {Sakamoto}, {Sarazin},
  {Schady}, {Stamatikos}, \& {Woosley}}]{2006Natur.444.1044G}
{Gehrels}, N., {Norris}, J.~P., {Barthelmy}, S.~D., {et~al.} 2006, Nature, 444,
  1044

\bibitem[{{Ghirlanda} {et~al.}(2010){Ghirlanda}, {Ghisellini}, \&
  {Nava}}]{2010A&A...510L...7G}
{Ghirlanda}, G., {Ghisellini}, G., \& {Nava}, L. 2010, Astron. Astrophys., 510,
  L7

\bibitem[{{Ghirlanda} {et~al.}(2016){Ghirlanda}, {Salafia}, {Pescalli},
  {Ghisellini}, {Salvaterra}, {Chassande-Mottin}, {Colpi}, {Nappo}, {D'Avanzo},
  {Melandri}, {Bernardini}, {Branchesi}, {Campana}, {Ciolfi}, {Covino},
  {G{\"o}tz}, {Vergani}, {Zennaro}, \& {Tagliaferri}}]{2016A&A...594A..84G}
{Ghirlanda}, G., {Salafia}, O.~S., {Pescalli}, A., {et~al.} 2016, Astron.
  Astrophys., 594, A84

\bibitem[{{G{\"o}tz} {et~al.}(2013){G{\"o}tz}, {Covino}, {Fern{\'a}ndez-Soto},
  {Laurent}, \& {Bo{\v s}njak}}]{2013MNRAS.431.3550G}
{G{\"o}tz}, D., {Covino}, S., {Fern{\'a}ndez-Soto}, A., {Laurent}, P., \&
  {Bo{\v s}njak}, {\v Z}. 2013, Mon. Not. R. Soc., 431, 3550

\bibitem[{{G{\"o}tz} {et~al.}(2014){G{\"o}tz}, {Laurent}, {Antier}, {Covino},
  {D'Avanzo}, {D'Elia}, \& {Melandri}}]{2014MNRAS.444.2776G}
{G{\"o}tz}, D., {Laurent}, P., {Antier}, S., {et~al.} 2014, Mon. Not. R. Soc.,
  444, 2776

\bibitem[{{Guiriec} {et~al.}(2010){Guiriec}, {Briggs}, {Connaugthon}, {Kara},
  {Daigne}, {Kouveliotou}, {van der Horst}, {Paciesas}, {Meegan}, {Bhat},
  {Foley}, {Bissaldi}, {Burgess}, {Chaplin}, {Diehl}, {Fishman}, {Gibby},
  {Giles}, {Goldstein}, {Greiner}, {Gruber}, {von Kienlin}, {Kippen},
  {McBreen}, {Preece}, {Rau}, {Tierney}, \&
  {Wilson-Hodge}}]{2010ApJ...725..225G}
{Guiriec}, S., {Briggs}, M.~S., {Connaugthon}, V., {et~al.} 2010, Astrophys.
  J., 725, 225

\bibitem[{{Guiriec} {et~al.}(2013){Guiriec}, {Daigne}, {Hasco{\"e}t},
  {Vianello}, {Ryde}, {Mochkovitch}, {Kouveliotou}, {Xiong}, {Bhat}, {Foley},
  {Gruber}, {Burgess}, {McGlynn}, {McEnery}, \&
  {Gehrels}}]{2013ApJ...770...32G}
{Guiriec}, S., {Daigne}, F., {Hasco{\"e}t}, R., {et~al.} 2013, Astrophys. J.,
  770, 32

\bibitem[{{Hakkila} {et~al.}(2008){Hakkila}, {Giblin}, {Norris}, {Fragile}, \&
  {Bonnell}}]{2008ApJ...677L..81H}
{Hakkila}, J., {Giblin}, T.~W., {Norris}, J.~P., {Fragile}, P.~C., \&
  {Bonnell}, J.~T. 2008, Astrophys. J. Lett., 677, L81

\bibitem[{{H.E.S.S.~Collaboration} {et~al.}(2011){H.E.S.S.~Collaboration},
  {Abramowski}, {Acero}, {Aharonian}, {Akhperjanian}, {Anton}, {Barnacka},
  {Barres de Almeida}, {Bazer-Bachi}, {Becherini}, {Becker}, {Behera},
  {Bernl{\"o}hr}, {Bochow}, {Boisson}, {Bolmont}, {Bordas}, {Borrel},
  {Brucker}, {Brun}, {Brun}, {B{\"u}hler}, {Bulik}, {B{\"u}sching}, {Carrigan},
  {Casanova}, {Cerruti}, {Chadwick}, {Charbonnier}, {Chaves}, {Cheesebrough},
  {Chounet}, {Clapson}, {Coignet}, {Conrad}, {Dalton}, {Daniel}, {Davids},
  {Degrange}, {Deil}, {Dickinson}, {Djannati-Ata{\"i}}, {Domainko}, {Drury},
  {Dubois}, {Dubus}, {Dyks}, {Dyrda}, {Egberts}, {Eger}, {Espigat}, {Fallon},
  {Farnier}, {Fegan}, {Feinstein}, {Fernandes}, {Fiasson}, {Fontaine},
  {F{\"o}rster}, {F{\"u}{\ss}ling}, {Gabici}, {Gallant}, {Gast}, {G{\'e}rard},
  {Gerbig}, {Giebels}, {Glicenstein}, {Gl{\"u}ck}, {Goret}, {G{\"o}ring},
  {Hague}, {Hampf}, {Hauser}, {Heinz}, {Heinzelmann}, {Henri}, {Hermann},
  {Hinton}, {Hoffmann}, {Hofmann}, {Hofverberg}, {Horns}, {Jacholkowska}, {de
  Jager}, {Jahn}, {Jamrozy}, {Jung}, {Kastendieck}, {Katarzy{\'n}ski}, {Katz},
  {Kaufmann}, {Keogh}, {Kerschhaggl}, {Khangulyan}, {Kh{\'e}lifi}, {Klochkov},
  {Klu{\'z}niak}, {Kneiske}, {Komin}, {Kosack}, {Kossakowski}, {Laffon},
  {Lamanna}, {Lenain}, {Lennarz}, {Lohse}, {Lopatin}, {Lu}, {Marandon},
  {Marcowith}, {Masbou}, {Maurin}, {Maxted}, {McComb}, {Medina}, {M{\'e}hault},
  {Moderski}, {Moulin}, {Naumann}, {Naumann-Godo}, {de Naurois}, {Nedbal},
  {Nekrassov}, {Nguyen}, {Nicholas}, {Niemiec}, {Nolan}, {Ohm}, {Olive}, {de
  O{\~n}a Wilhelmi}, {Opitz}, {Ostrowski}, {Panter}, {Paz Arribas},
  {Pedaletti}, {Pelletier}, {Petrucci}, {Pita}, {P{\"u}hlhofer}, {Punch},
  {Quirrenbach}, {Raue}, {Rayner}, {Reimer}, {Reimer}, {Renaud}, {de Los
  Reyes}, {Rieger}, {Ripken}, {Rob}, {Rosier-Lees}, {Rowell}, {Rudak},
  {Rulten}, {Ruppel}, {Ryde}, {Sahakian}, {Santangelo}, {Schlickeiser},
  {Sch{\"o}ck}, {Sch{\"o}nwald}, {Schwanke}, {Schwarzburg}, {Schwemmer},
  {Shalchi}, {Sikora}, {Skilton}, {Sol}, {Spengler}, {Stawarz}, {Steenkamp},
  {Stegmann}, {Stinzing}, {Sushch}, {Szostek}, {Tam}, {Tavernet}, {Terrier},
  {Tibolla}, {Tluczykont}, {Valerius}, {van Eldik}, {Vasileiadis}, {Venter},
  {Vialle}, {Viana}, {Vincent}, {Vivier}, {V{\"o}lk}, {Volpe}, {Vorobiov},
  {Vorster}, {Wagner}, {Ward}, {Wierzcholska}, {Zajczyk}, {Zdziarski}, {Zech},
  {Zechlin}, \& {H.E.S.S.~Collaboration}}]{2011APh....34..738H}
{H.E.S.S.~Collaboration}, {Abramowski}, A., {Acero}, F., {et~al.} 2011,
  Astroparticle Physics, 34, 738

\bibitem[{{Ioka} \& {Nakamura}(2001)}]{2001ApJ...554L.163I}
{Ioka}, K. \& {Nakamura}, T. 2001, Astrophys. J. Lett., 554, L163

\bibitem[{{Jacob} \& {Piran}(2008)}]{2008JCAP...01..031J}
{Jacob}, U. \& {Piran}, T. 2008, JCAP, 1, 031

\bibitem[{{Kocevski} \& {Liang}(2003)}]{2003ApJ...594..385K}
{Kocevski}, D. \& {Liang}, E. 2003, Astrophys. J., 594, 385

\bibitem[{{Laurent} {et~al.}(2011){Laurent}, {G{\"o}tz}, {Bin{\'e}truy},
  {Covino}, \& {Fernandez-Soto}}]{2011PhRvD..83l1301L}
{Laurent}, P., {G{\"o}tz}, D., {Bin{\'e}truy}, P., {Covino}, S., \&
  {Fernandez-Soto}, A. 2011, Phys.Rev. D, 83, 121301

\bibitem[{{Lin} {et~al.}(2016){Lin}, {Li}, \& {Chang}}]{2016MNRAS.463..375L}
{Lin}, H.-N., {Li}, X., \& {Chang}, Z. 2016, Mon. Not. R. Soc., 463, 375

\bibitem[{{Lu} {et~al.}(2006){Lu}, {Qin}, {Zhang}, \&
  {Yi}}]{2006MNRAS.367..275L}
{Lu}, R.-J., {Qin}, Y.-P., {Zhang}, Z.-B., \& {Yi}, T.-F. 2006, Mon. Not. R.
  Soc., 367, 275

\bibitem[{{MAGIC Collaboration} {et~al.}(2008){MAGIC Collaboration}, {Albert},
  {Aliu}, {Anderhub}, {Antonelli}, {Antoranz}, {Backes}, {Baixeras}, {Barrio},
  {Bartko}, {Bastieri}, {Becker}, {Bednarek}, {Berger}, {Bernardini},
  {Bigongiari}, {Biland}, {Bock}, {Bonnoli}, {Bordas}, {Bosch-Ramon}, {Bretz},
  {Britvitch}, {Camara}, {Carmona}, {Chilingarian}, {Commichau}, {Contreras},
  {Cortina}, {Costado}, {Covino}, {Curtef}, {Dazzi}, {de Angelis}, {de Cea Del
  Pozo}, {Delgado Mendez}, {de Los Reyes}, {de Lotto}, {de Maria}, {de Sabata},
  {Dominguez}, {Dorner}, {Doro}, {Errando}, {Fagiolini}, {Ferenc},
  {Fern{\'a}ndez}, {Firpo}, {Fonseca}, {Font}, {Galante}, {Garc{\'{\i}}a
  L{\'o}pez}, {Garczarczyk}, {Gaug}, {Goebel}, {Hayashida}, {Herrero},
  {H{\"o}hne}, {Hose}, {Hsu}, {Huber}, {Jogler}, {Kranich}, {La Barbera},
  {Laille}, {Leonardo}, {Lindfors}, {Lombardi}, {Longo}, {L{\'o}pez}, {Lorenz},
  {Majumdar}, {Maneva}, {Mankuzhiyil}, {Mannheim}, {Maraschi}, {Mariotti},
  {Mart{\'{\i}}nez}, {Mazin}, {Meucci}, {Meyer}, {Miranda}, {Mirzoyan},
  {Moles}, {Moralejo}, {Nieto}, {Nilsson}, {Ninkovic}, {Otte}, {Oya},
  {Panniello}, {Paoletti}, {Paredes}, {Pasanen}, {Pascoli}, {Pauss}, {Pegna},
  {Perez-Torres}, {Persic}, {Peruzzo}, {Piccioli}, {Prada}, {Prandini},
  {Puchades}, {Raymers}, {Rhode}, {Rib{\'o}}, {Rico}, {Rissi}, {Robert},
  {R{\"u}gamer}, {Saggion}, {Saito}, {Salvati}, {Sanchez-Conde}, {Sartori},
  {Satalecka}, {Scalzotto}, {Scapin}, {Schmitt}, {Schweizer}, {Shayduk},
  {Shinozaki}, {Sidro}, {Sierpowska-Bartosik}, {Sillanp{\"a}{\"a}},
  {Sobczynska}, {Spanier}, {Stamerra}, {Stark}, {Takalo}, {Tavecchio},
  {Temnikov}, {Tescaro}, {Teshima}, {Tluczykont}, {Torres}, {Turini}, {Vankov},
  {Venturini}, {Vitale}, {Wagner}, {Wittek}, {Zabalza}, {Zandanel}, {Zanin},
  {Zapatero}, {Ellis}, {Mavromatos}, {Nanopoulos}, {Sakharov}, \&
  {Sarkisyan}}]{2008PhLB..668..253M}
{MAGIC Collaboration}, {Albert}, J., {Aliu}, E., {et~al.} 2008, Physics Letters
  B, 668, 253

\bibitem[{{Malesani} {et~al.}(2015){Malesani}, {Kruehler}, {Xu}, {Pugliese},
  {Watson}, {Fynbo}, {Milvang-Jensen}, {de Ugarte Postigo}, {Tanvir}, {D'Elia},
  {Wiersema}, {Greiner}, \& {Japelj}}]{2015GCN..17755...1M}
{Malesani}, D., {Kruehler}, T., {Xu}, D., {et~al.} 2015, GRB Coordinates
  Network, 17755

\bibitem[{{Mattingly}(2005)}]{2005LRR.....8....5M}
{Mattingly}, D. 2005, Living Reviews in Relativity, 8, 5

\bibitem[{{Mochkovitch} {et~al.}(2016){Mochkovitch}, {Heussaff}, {Atteia},
  {Bo{\c c}i}, \& {Hafizi}}]{2016A&A...592A..95M}
{Mochkovitch}, R., {Heussaff}, V., {Atteia}, J.~L., {Bo{\c c}i}, S., \&
  {Hafizi}, M. 2016, Astron. Astrophys., 592, A95

\bibitem[{{Norris}(2002)}]{2002ApJ...579..386N}
{Norris}, J.~P. 2002, Astrophys. J., 579, 386

\bibitem[{{Norris} \& {Bonnell}(2006)}]{2006ApJ...643..266N}
{Norris}, J.~P. \& {Bonnell}, J.~T. 2006, Astrophys. J., 643, 266

\bibitem[{{Norris} {et~al.}(2000){Norris}, {Marani}, \&
  {Bonnell}}]{2000ApJ...534..248N}
{Norris}, J.~P., {Marani}, G.~F., \& {Bonnell}, J.~T. 2000, Astrophys. J., 534,
  248

\bibitem[{{Norris} {et~al.}(2001){Norris}, {Scargle}, \&
  {Bonnell}}]{2001grba.conf...40N}
{Norris}, J.~P., {Scargle}, J.~D., \& {Bonnell}, J.~T. 2001, in Gamma-ray
  Bursts in the Afterglow Era, ed. E.~{Costa}, F.~{Frontera}, \& J.~{Hjorth},
  40

\bibitem[{{Peng} {et~al.}(2011){Peng}, {Yin}, {Bi}, {Bao}, \&
  {Ma}}]{2011AN....332...92P}
{Peng}, Z.~Y., {Yin}, Y., {Bi}, X.~W., {Bao}, Y.~Y., \& {Ma}, L. 2011,
  Astronomische Nachrichten, 332, 92

\bibitem[{{Peterson} {et~al.}(1998){Peterson}, {Wanders}, {Horne}, {Collier},
  {Alexander}, {Kaspi}, \& {Maoz}}]{1998PASP..110..660P}
{Peterson}, B.~M., {Wanders}, I., {Horne}, K., {et~al.} 1998, Publications of
  the Astronomical Society of the Pacific, 110, 660

\bibitem[{{Planck Collaboration} {et~al.}(2016){Planck Collaboration}, {Ade},
  {Aghanim}, {Arnaud}, {Ashdown}, {Aumont}, {Baccigalupi}, {Banday},
  {Barreiro}, {Bartlett}, \& et~al.}]{2016A&A...594A..13P}
{Planck Collaboration}, {Ade}, P.~A.~R., {Aghanim}, N., {et~al.} 2016, Astron.
  Astrophys., 594, A13

\bibitem[{{Rodr{\'{\i}}guez Mart{\'{\i}}nez} \&
  {Piran}(2006)}]{2006JCAP...04..006R}
{Rodr{\'{\i}}guez Mart{\'{\i}}nez}, M. \& {Piran}, T. 2006, JCAP, 4, 006

\bibitem[{{Rodr{\'{\i}}guez Mart{\'{\i}}nez} {et~al.}(2006){Rodr{\'{\i}}guez
  Mart{\'{\i}}nez}, {Piran}, \& {Oren}}]{2006JCAP...05..017R}
{Rodr{\'{\i}}guez Mart{\'{\i}}nez}, M., {Piran}, T., \& {Oren}, Y. 2006, JCAP,
  5, 017

\bibitem[{{Ryde}(2005)}]{2005A&A...429..869R}
{Ryde}, F. 2005, Astron. Astrophys., 429, 869

\bibitem[{{Sakamoto} {et~al.}(2011){Sakamoto}, {Barthelmy}, {Baumgartner},
  {Cummings}, {Fenimore}, {Gehrels}, {Krimm}, {Markwardt}, {Palmer}, {Parsons},
  {Sato}, {Stamatikos}, {Tueller}, {Ukwatta}, \& {Zhang}}]{2011ApJS..195....2S}
{Sakamoto}, T., {Barthelmy}, S.~D., {Baumgartner}, W.~H., {et~al.} 2011,
  Astrophys. J. Supp., 195, 2

\bibitem[{{Salmonson}(2000)}]{2000ApJ...544L.115S}
{Salmonson}, J.~D. 2000, Astrophys. J. Lett., 544, L115

\bibitem[{{Salvaterra} {et~al.}(2012){Salvaterra}, {Campana}, {Vergani},
  {Covino}, {D'Avanzo}, {Fugazza}, {Ghirlanda}, {Ghisellini}, {Melandri},
  {Nava}, {Sbarufatti}, {Flores}, {Piranomonte}, \&
  {Tagliaferri}}]{2012ApJ...749...68S}
{Salvaterra}, R., {Campana}, S., {Vergani}, S.~D., {et~al.} 2012, Astrophys.
  J., 749, 68

\bibitem[{{Schaefer}(2007)}]{2007ApJ...660...16S}
{Schaefer}, B.~E. 2007, Astrophys. J., 660, 16

\bibitem[{{Selsing} {et~al.}(2017){Selsing}, {Kr{\"u}hler}, {Malesani},
  {D'Avanzo}, {Schulze}, {Palmerio}, {Vergani}, {Japelj}, {Milvang-Jensen},
  {Watson}, {Jakobsson}, {Bolmer}, {Cano}, {Covino}, {Christensen}, {D'Elia},
  {de Ugarte Postigo}, {Fynbo}, {Gomboc}, {Heintz}, {Kaper}, {Levan},
  {Piranomonte}, {Pugliese}, {S{\'a}nchez-Ram{\'{\i}}rez}, {Sparre}, {Tanvir},
  {Th{\"o}ne}, \& {Wiersema}}]{2017arXiv170701452S}
{Selsing}, J., {Kr{\"u}hler}, T., {Malesani}, D., {et~al.} 2017, ArXiv e-prints
  [\eprint[arXiv]{1707.01452}]

\bibitem[{{Selsing} {et~al.}(2016){Selsing}, {Vreeswijk}, {Japelj}, {D'Elia},
  {de Ugarte Postigo}, {Pugliese}, {Xu}, {Malesani}, {Kruehler}, \&
  {Fynbo}}]{2016GCN..19274...1S}
{Selsing}, J., {Vreeswijk}, P.~M., {Japelj}, J., {et~al.} 2016, GRB Coordinates
  Network, 19274

\bibitem[{{Shen} {et~al.}(2005){Shen}, {Song}, \& {Li}}]{2005MNRAS.362...59S}
{Shen}, R.-F., {Song}, L.-M., \& {Li}, Z. 2005, Mon. Not. R. Soc., 362, 59

\bibitem[{{Toma} {et~al.}(2012){Toma}, {Mukohyama}, {Yonetoku}, {Murakami},
  {Gunji}, {Mihara}, {Morihara}, {Sakashita}, {Takahashi}, {Wakashima},
  {Yonemochi}, \& {Toukairin}}]{2012PhRvL.109x1104T}
{Toma}, K., {Mukohyama}, S., {Yonetoku}, D., {et~al.} 2012, Physical Review
  Letters, 109, 241104

\bibitem[{{Ukwatta} {et~al.}(2012){Ukwatta}, {Dhuga}, {Stamatikos}, {Dermer},
  {Sakamoto}, {Sonbas}, {Parke}, {Maximon}, {Linnemann}, {Bhat}, {Eskandarian},
  {Gehrels}, {Abeysekara}, {Tollefson}, \& {Norris}}]{2012MNRAS.419..614U}
{Ukwatta}, T.~N., {Dhuga}, K.~S., {Stamatikos}, M., {et~al.} 2012, Mon. Not. R.
  Soc., 419, 614

\bibitem[{{Ukwatta} {et~al.}(2010){Ukwatta}, {Stamatikos}, {Dhuga}, {Sakamoto},
  {Barthelmy}, {Eskandarian}, {Gehrels}, {Maximon}, {Norris}, \&
  {Parke}}]{2010ApJ...711.1073U}
{Ukwatta}, T.~N., {Stamatikos}, M., {Dhuga}, K.~S., {et~al.} 2010, Astrophys.
  J., 711, 1073

\bibitem[{{Vasileiou} {et~al.}(2013){Vasileiou}, {Jacholkowska}, {Piron},
  {Bolmont}, {Couturier}, {Granot}, {Stecker}, {Cohen-Tanugi}, \&
  {Longo}}]{2013PhRvD..87l2001V}
{Vasileiou}, V., {Jacholkowska}, A., {Piron}, F., {et~al.} 2013, Phys.Rev. D,
  87, 122001

\bibitem[{{Zhang}(2006)}]{2006Natur.444.1010Z}
{Zhang}, B. 2006, Nature, 444, 1010

\end{thebibliography}
\end{document}